\documentclass[12pt]{article}
\usepackage{axodraw,epsfig}
\title{\bf The neutrino self-energy in a magnetized medium}
\bigskip
\author{Alberto  Bravo Garc\'{i}a$^1$,\,\,Kaushik Bhattacharya$^1$,\,\,Sarira
  Sahu$^{1,2}$ 
\thanks{e-mail
addresses:abravo@nucleares.unam.mx, kaushik@nucleares.unam.mx, 
sarira@nucleares.unam.mx 
}\\
\normalsize
$^1$Instituto de Ciencias Nucleares,\\
Universidad
Nacional Autonoma de Mexico, Circuito Exterior, C.U.,\\
A. Postal 70-543, C.P. 04510 Mexico DF, Mexico\\
$^2$Institute of Astrophysics and Particle Physics
and Department of Physics\\
National Taiwan University, Taipei 106, Taiwan\\
\normalsize 
}
\begin{document}
\date{ }
\maketitle
\begin{abstract}
In this work we calculate the neutrino self-energy in presence of a
magnetized medium. The magnetized medium consists of electrons,
positrons, neutrinos and a uniform classical magnetic field. The
calculation is done assuming the background magnetic field is weak
compared to the $W$-Boson mass squared, as a consequence of which only
linear order corrections in the field are included in the $W$ boson
propagator. The electron propagator consists all order corrections in
the background field. Although the neutrino self-energy in a
magnetized medium in various limiting cases has been calculated previously
in this article we produce the most general expression of the
self-energy in absence of the Landau quantization of the charged gauge
fields. We calculate the effect of the Landau quantization of the
charged leptons on the neutrino self-energy in the general case. Our
calculation is specifically suited for situations where the background
plasma may be CP symmetric.
\end{abstract}
\section{Introduction}
\label{intro}
The topic of neutrino self-energy in a thermal medium or a magnetized
medium has attracted much attention in the last two decades
\cite{Elmfors:1996gy, D'Olivo:1992vm, D'Olivo:1989cr}. Presence of
magnetic field in active galactic nuclei as well as accretion disk of
merging objects \cite{Donati:2005tw} and progenitors of Gamma Ray
Bursts (GRBs)\cite{Granot:2003hx, Li:2006ft} are obvious. So it is
important to study the combined effect of both matter and magnetic
field on neutrino propagation.  As neutrinos are produced in
elementary particle reactions, their presence in all of the above
mentioned objects is a fact and once produced they propagate through
the medium containing charged leptons, nucleons and/or neutrinos in
presence of possible high/low magnetic fields.

In the present article we assume the magnetic field to be less than
the critical field corresponding the $W^{\pm}$ bosons and consequently
we take only linear order, in the magnetic field strength, corrections
to its propagator. The gauge bosons are assumed to be not in thermal
equilibrium and so their thermal modifications are not used.  In the
present case we have worked in the unitary gauge and have not discussed
about the gauge independence of the result, as it is noted that in
such calculations the self-energy generally is dependent on the gauge
choice but the dispersion relation is independent of the gauge
\cite{Erdas:1990gy, Erdas:2000iq}.

In this context we mention some differences from previous works namely
those in \cite{Erdas:1990gy,Erdas:2000iq}. In most of the earlier
works the authors have chosen a particular gauge as where the gauge
parameter is unity, the Feynman gauge, and then tried to show that the
dispersion relations are independent of the gauge choice. While in
\cite{Elmfors:1996gy} the authors worked in unitary gauge to order
$G_F^2$ but neither they nor the authors of
Ref.\cite{Erdas:1990gy,Erdas:2000iq} did transparently exhibit the
transverse and longitudinal decomposition of the neutrino self-energy
which is expected in the presence of a magnetic field.  In a related
work \cite{Elizalde:2004mw} the authors decomposed the self-energy in
the transverse and longitudinal parts and calculated the neutrino
self-energy in a magnetized medium for high magnetic fields. In the
last reference the authors approximated the electron propagator by its
lowest Landau level value while used the most general $W$-boson
propagator as they assumed that $M_W^2 \gg e{\mathcal B} \gg m^2_e$,
where ${\mathcal B}$ is the magnitude of the magnetic field and $m_e$,
$M_W$ are the electron and the $W$ boson masses.  In
ref.~\cite{Esposito:1995db} the propagation of neutrino in an
isotropic magnetized medium has been studied where they consider the
magnetic field to be weak compared to the electron mass $e{\mathcal B}
\ll m^2_e$.  In the present work we use the unitary gauge to calculate
the self-energy of the neutrino to order $G^2_F$ and we find that the
self-energy expression shows transverse and longitudinal neutrino
momentum dependence as in \cite{Elizalde:2004mw} but we do not assume
$e{\mathcal B} \gg m^2_e$ and consequently we use the full electron
propagator in presence of the magnetized medium while for the $W$
boson propagator we only take linear order corrections in the magnetic
field as in our case $M_W^2 \gg e{\mathcal B}$. In the present article
we only aasume that $T,\,\mu_\ell,\,\sqrt{\mathcal B} < M_W$, where
$T$ and $\mu$ are the temperature of the heat bath and the chemical
potential of the charged leptons. We recover all the relevant result in the
limit when ${\mathcal B} \to 0$ and matches with the results in
\cite{notr}.

The paper is organized as follows. In section \ref{renexp} we discuss
about the general properties of the neutrino self-energy in a
magnetized medium and its possible form. The various diagrams
contributing to the neutrino self-energy and their individual
contributions are also written down in the next section.  In section
\ref{contrib} we calculate the contributions from the different
diagrams and discuss about our scheme of computation. Section
\ref{dispr} summarizes the results of the previous section and there
we write the dispersion relation of the neutrino in a magnetized
medium. Finally we summarize our results in section \ref{conclu}.  
\section{General expression for the neutrino self-energy in a magnetized 
medium}
\label{renexp}
The most general form of neutrino-self energy in presence of a
magnetized medium can be written as:
\begin{eqnarray}
\Sigma (k) &=& R \Big( a_\parallel k^\mu_\parallel + a_\perp k^\mu_\perp + b u^\mu 
+ c b^\mu \Big) \gamma_\mu L \,,
\label{sigmanb}
\end{eqnarray}
where $k^\mu_\parallel=(k^0, k^3)$ and $k^\mu_\perp=(k^1,
k^2)$. $u^\mu$ stands for the 4-velocity of the centre-of-mass of the
medium which looks like:
\begin{eqnarray}
u^\mu = (1, {\bf 0})\,,
\label{u}
\end{eqnarray}
in the rest frame of the medium. The $u^\mu$ is normalized in such a way that,
\begin{eqnarray}
u^\mu u_\mu =1\,.
\label{unorm}
\end{eqnarray} 
Likewise the effect of the magnetic field enters through the 4-vector
$b^\mu$ which is defined in such a way that the frame in which the
medium is at rest,
\begin{eqnarray}
b^\mu = (0, \hat{\bf b})\,,
\label{b}
\end{eqnarray}
where we denote the magnetic field vector by ${\mathcal B} \hat{\bf
b}$. The 4-vector $b^\mu$ is defined in such a way that,
\begin{eqnarray} 
b^\mu b_\mu = -1\,.
\label{bnorm}
\end{eqnarray}
In this article we take the background classical magnetic field vector
to be along the $z$-axis and consequently $b^\mu=(0,0,0,1)$.  The
projection operators are conventionally defined as $R=
\frac12(1+\gamma_5)$ and $L = \frac12(1-\gamma_5)$.  Calculting the
neutrino dispersion relation from Eq.~(\ref{sigmanb}) we get:
\begin{eqnarray}
(1-a_\parallel)\omega_{_\ell} = \pm \left[\left((1-a_\parallel)k_3 + c\right)^2
+ (1-a_\perp)^2 k_\perp^2\right]^{1/2} + b\,.
\label{mndisp}
\end{eqnarray}
In deriving the above dispersion relation Eq.~(\ref{unorm}) and
Eq.~(\ref{bnorm}) has been used and we have dropped terms like $b\cdot
u$, $k_\perp \cdot u$, $k_\perp \cdot b$ which are zero in the rest
frame of the medium, and $k_\perp^2=k_1^2 + k_2^2$. 

In the unitary gauge the three diagrams corresponding to the neutrino
self-energy are as given in Fig.~\ref{f:selfen1} and
Fig.~\ref{f:selfen2}.
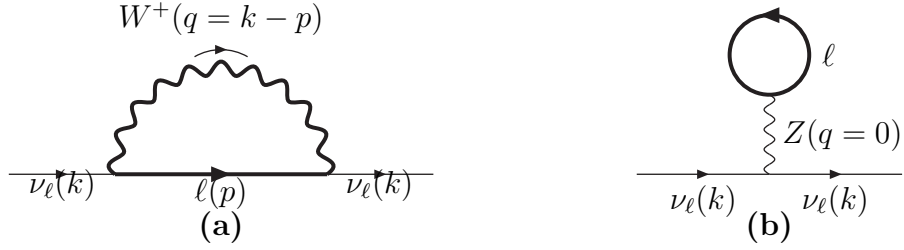
\begin{figure}[h!]
\begin{center}
\begin{picture}(180,75)(-90,-20)
\ArrowLine(40,0)(80,0)
\Text(60,-10)[b]{$\nu_\ell(k)$}
\Text(0,-13)[b]{$\ell(p)$}
\ArrowLine(-80,0)(-40,0)
\Text(-60,-10)[b]{$\nu_\ell(k)$}
\ArrowArcn(0,27)(20,120,60)
\SetWidth{1.5}
\PhotonArc(0,0)(40,0,180){2.5}{10.5}
\ArrowLine(-40,0)(40,0)
\Text(0,53)[b]{$W^+(q=k-p)$}
\Text(0,-20)[]{\bf (a)}
\end{picture}
\qquad
\begin{picture}(180,75)(-90,-20)
\ArrowLine(0,0)(50,0)
\Text(25,-10)[c]{$\nu_\ell(k)$}
\ArrowLine(-50,0)(0,0)
\Text(-25,-10)[c]{$\nu_\ell(k)$}
\Photon(0,0)(0,30)24
\Text(5,15)[l]{$Z(q=0)$}
\Text(20,45)[l]{$\ell$}
\SetWidth{1.5}
\ArrowArc(0,45)(15,-90,270)
\Text(0,-20)[]{\bf (b)}
\end{picture}
\caption[]{\sf One-loop diagrams for neutrino self-energy in a
magnetized medium. Diagram (a) is the $W$ exchange diagram and diagram
(b) is the tadpole diagram. The heavy internal lines represent the $W$
and the charged lepton propagators in a magnetized medium.
\label{f:selfen1}}
\end{center}
\end{figure}
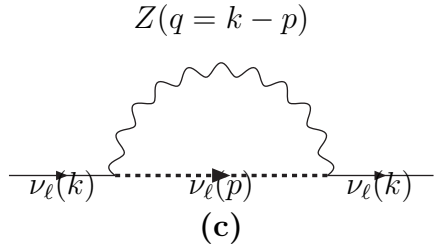
\begin{figure}[h!]
\begin{center}
\begin{picture}(180,75)(-90,-20)
\ArrowLine(40,0)(80,0)
\Text(60,-10)[b]{$\nu_\ell(k)$}
\ArrowLine(-80,0)(-40,0)
\Text(0,-10)[b]{$\nu_\ell(p)$}
\Text(-60,-10)[b]{$\nu_\ell(k)$}
\PhotonArc(0,0)(40,0,180){2.5}{10.5}
\Text(0,53)[b]{$Z(q=k-p)$}
\Text(0,-20)[]{\bf (c)}
\SetWidth{1.3}
\DashArrowLine(-40,0)(40,0){2}
\end{picture}
\caption[]{\sf The One-loop diagram for neutrino self-energy in a
magnetized medium. This diagram is for the $Z$ exchange. The heavy
dashed internal neutrino line corresponds to the neutrino propagator in a
thermal medium.
\label{f:selfen2}}
\end{center}
\end{figure}
The one-loop neutrino self-energy in a magnetized medium is comprised
of three pieces, one coming from the $W$-exchange diagram which we
will call $\Sigma^W (k)$, one from the tadpole diagram which we will
designate by $\Sigma^T (k)$ and one from the $Z$-exchange diagram
which will be denoted by $\Sigma^Z (k)$. The total self-energy of the neutrino in a magnetized medium then becomes:
\begin{eqnarray}
\Sigma(k) = \Sigma^W(k) + \Sigma^T (k) + \Sigma^Z (k)\,.
\label{tsen}
\end{eqnarray}
Each of the individual terms appearing in the right-hand side of the
above equation can be expressed as in Eq.~(\ref{sigmanb}). 
Expressing Eq.~(\ref{tsen}) as in Eq.~(\ref{sigmanb}) gives,
\begin{eqnarray}
a &=& a_W + a_T + a_Z\,,
\label{at}\\
b &=& b_W + b_T + b_Z\,,
\label{bt}\\
c &=& c_W + c_T + c_Z\,.
\label{ct}
\end{eqnarray}
In Eq.~(\ref{at}) $a_W$, $a_T$ and $a_Z$ are composed of the
parallel and perpendicular parts as shown in Eq.~(\ref{sigmanb}).
\section{Contribution from the various diagrams}
\label{contrib}
The individual terms on the right hand side of Eq.~(\ref{tsen}) can be
explicitly written as:
\begin{eqnarray}
-i\Sigma^W(k)=\int\frac{d^4 p}{(2\pi)^4}\left(\frac{-ig}{\sqrt{2}}\right)
\gamma_\mu\, L \,iS_{\ell}(p)\left(\frac{-ig}{\sqrt{2}}\right)\gamma_\nu 
 \,L\,i W^{\mu \nu}(q)\,,
\label{wexch}
\end{eqnarray}
\begin{eqnarray}
-i\Sigma^T(k)= -\left(\frac{g}{2\cos \theta_W}\right)^2 R\,
\gamma_\mu\,iZ^{\mu \nu}(0)\int\frac{d^4 p}{(2\pi)^4} {\rm Tr}
\left[\gamma_\nu \,(c_V + c_A \gamma_5)\,iS_{\ell}(p)\right]\,,   
\label{tad}
\end{eqnarray}
and
\begin{eqnarray}
-i\Sigma^Z(k)=\int\frac{d^4 p}{(2\pi)^4}\left(\frac{-ig}
{\sqrt{2}\cos\theta_W}\right)
\gamma_\mu\, L \,iS_{\nu_\ell}(p)\left(\frac{-ig}{\sqrt{2}\cos\theta_W}\right)
\gamma_\nu\,L\,i Z^{\mu \nu}(q)\,.
\label{Zexch}
\end{eqnarray}
In the above expressions $g$ is the weak coupling constant and
$\theta_W$ is the Weinberg angle.  The quantities $c_V$ and $c_A$ are
the vector and axial-vector coupling constants
which come in the neutral-current interaction of
electrons, protons ($p$), neutrons ($n$) and neutrinos with the $Z$
boson.  Their forms are as follows,
\begin{eqnarray}
c_V=\left \{\begin{array}
{r@{\quad\quad}l}
-\frac{1}{2}+2\sin^2\theta_W & e\\
\frac{1}{2} & {\nu_e}\\ \frac{1}{2}-2\sin^2\theta_W & {{p}}\\
-\frac{1}{2} & {{n}}
\end{array}\right.,
\label{cv}
\end{eqnarray}
and
\begin{eqnarray}
c_A=\left \{\begin{array}
{r@{\quad\quad}l}
-\frac{1}{2} & {\nu},{{p}}\\
\frac{1}{2} & e,{{n}}
\end{array}\right..
\label{ca}
\end{eqnarray}
Here $W^{\mu \nu}(q)$ and $S_\ell(p)$ stand for the $W$-boson propagator
and charged lepton propagator respectively in
presence of a magnetized plasma. The $Z^{\mu \nu}(q)$ is the $Z$-boson
propagator in vacuum and $S_{\nu_\ell}(p)$ is the neutrino propagator
in a thermal bath of neutrinos.  The form of the charged lepton
propagator in a magnetized medium is given by
\cite{Bhattacharya:2002aj, Bhattacharya:2004nj}:
\begin{eqnarray}
S_\ell (p) = S^0_\ell(p) + S^\beta_\ell (p)\,,
\label{slp} 
\end{eqnarray}
where $S^0_\ell(p)$ and $S^\beta_\ell (p)$ are the charged lepton
propagators in presence of an uniform background magnetic field and in
a magnetized medium respectively. In this article we will always
assume that the magnetic field is directed towards the $z$-axis of the
coordinate system. With this choice we have \cite{D'Olivo:2002sp}:
\begin{eqnarray}
i S^0_\ell(p) = \int_0^\infty e^{\phi(p,s)} G(p,s)\,ds\,,
\label{sl0p}
\end{eqnarray}
where,
\begin{eqnarray}
\phi(p,s) = is(p^2_\parallel - m_\ell^2 - \frac{\tan z}{z} p^2_\perp)\,.
\label{phasel}
\end{eqnarray}
In the above expression 
\begin{eqnarray}
p^2_\parallel &=& p_0^2 - p_3^2\,,\\ 
p^2_\perp &=& p_1^2 + p_2^2\,, 
\end{eqnarray}
and $z= e{\mathcal B}s$ where $e$ is the magnitude of
the electron charge, ${\mathcal B}$ is the magnitude of the magnetic
field and $m_\ell$ is the mass of the charged lepton. In the above
equation we have not written another contribution to the phase which
is $\epsilon |s|$ where $\epsilon$ is an infinitesimal quantity. This
term renders the $s$ integration convergent. We do not explicitly
write this term but implicitly we assume the existence of it and it
will be written if required.  The above equation can also be written
as:
\begin{eqnarray}
\phi(p,s) = \psi(p_0) - is[p^2_3 + \frac{\tan z}{z} p^2_\perp]\,,
\label{phaselu}
\end{eqnarray}
where,
\begin{eqnarray}
\psi(p_0)=is(p_0^2 - m_\ell^2)\,.
\label{psi}
\end{eqnarray}
The other term in Eq.~(\ref{sl0p}) is given as:
\begin{eqnarray}
G(p,s)=\sec^2 z \left[{\rlap A/} + i {\rlap B/} \gamma_5 +m_\ell(\cos^2 z - 
i \Sigma^3 \sin z \cos z)\right]\,,
\label{gps}
\end{eqnarray}
where,
\begin{eqnarray}
A_\mu &=& p_\mu -\sin^2 z (p\cdot u\,\, u_\mu - p\cdot b \,\,b_\mu)\,,
\label{A}\\
B_\mu &=& \sin z\cos z (p\cdot u \,\,b_\mu - p\cdot b \,\,u_\mu)\,, 
\label{B}
\end{eqnarray}
and 
\begin{eqnarray}
\Sigma^3 = \gamma_5 {\rlap /b} {\rlap /u}\,. 
\label{sigm3}
\end{eqnarray}
The second term on the right-hand side of Eq.~(\ref{slp}) denotes the
medium contribution to the charged lepton propagator and its form is
given by:
\begin{eqnarray}
S^\beta_\ell(p) = i \eta_F(p\cdot u)\int_{-\infty}^\infty e^{\phi(p,s)} G(p,s)
\,ds\,,
\label{slbp}
\end{eqnarray}
where $\eta_F(p\cdot u)$ contains the distribution functions of the
particles in the medium and its form is:
\begin{eqnarray}
\eta_F (p \cdot u) = \frac{\theta(p\cdot u)}{e^{\beta(p\cdot u - 
\mu_\ell)} + 1 } +
\frac{\theta(- p\cdot u)}{e^{-\beta(p\cdot u - \mu_\ell)} + 1}\,,
\label{eta} 
\end{eqnarray}
where $\beta$ and $\mu_\ell$ are the inverse of the medium temperature
and the chemical potential of the charged lepton.

The form of the $W$-propagator in presence of a uniform magnetic field
along the $z$-direction is presented in \cite{Erdas:1998uu} and in
this article we only use the linearized (in the magnetic field) form
of it. The reason we assume a linearized form of the $W$-propagator is
because the magnitude of the magnetic field we consider is such that
$e{\mathcal B} \ll {M^2_W}$. In this limit the form of the
propagator is:
\begin{eqnarray}
W^{\mu \nu}(q) &=& -\frac{1}{q^2 - M^2_W}\left[ g^{\mu \nu} - \frac{1}{M^2_W}
(q^\mu q^\nu + \frac{ie}{2}F^{\mu \nu})\right] + 
\frac{2ie F^{\mu \nu}}{(q^2 - M^2_W)^2}\nonumber\\
&-& \frac{1}{(q^2 - \frac{M_W^2}{\xi})}\frac{1}{M_W^2}
\left(q^\mu q^\nu + \frac{ie}{2}F^{\mu \nu}\right)\,.
\label{wmunu}
\end{eqnarray}
In the above equation $M_W$ is the $W$-boson mass and $\xi$ is the
gauge-parameter.  In the unitary gauge $\xi=0$ and the last term on
the right-hand side of the above equation drops out.  More over in our
work we assume $q^2 \ll M_W^2$ and keep terms up to $1/M_W^4$
in the $W$ propagator. The propagator form in the unitary gauge and in
low momentum limit then looks like:
\begin{eqnarray}
W^{\mu \nu}(q) = \frac{g^{\mu \nu}}{M^2_W}\left(1 + \frac{q^2}{M^2_W}\right)
- \frac{q^\mu q^\nu}{M^4_W} + \frac{3ie}{2 M^4_W} F^{\mu \nu}\,.
\label{impw}
\end{eqnarray}
%
\subsection{The $W$-exchange diagram}
\label{wexchd}
The $W$-exchange contribution to the neutrino self-energy is given in
Eq.~(\ref{wexch}). In this subsection we will calculate the
contribution to the neutrino self-energy coming from the magnetized
plasma of electrons and positrons.  With the propagator of the $W$
boson as given in Eq.~(\ref{impw}) and the medium contribution from
the lepton propagator in Eq.~(\ref{slp}) we can write
Eq.~(\ref{wexch}) in a magnetized medium. This form of the self-energy
can be broken up into three parts as:
\begin{eqnarray}
-i\Sigma^W_1(k) &=& \left(\frac{g}{\sqrt{2}}\right)^2 \frac{1}{M^2_W}\int
\frac{d^4 p}{(2\pi)^4} R\,
\gamma_\mu\,S^\beta_{\ell}(p)\gamma_\nu
\,L\, g^{\mu \nu} \left(1 + \frac{q^2}{M^2_W}\right)\,,
\label{w1}\\
-i\Sigma^W_2(k) &=& -\left(\frac{g}{\sqrt{2}}\right)^2 \frac{1}{M^4_W}\int
\frac{d^4 p}{(2\pi)^4} R\,
\gamma_\mu\,S^\beta_{\ell}(p)\gamma_\nu
\,L\, q^\mu q^\nu\,,
\label{w2}\\
-i\Sigma^W_3(k) &=&  \frac{3ie}{2 M^4_W} F^{\mu \nu}
\left(\frac{g}{\sqrt{2}}\right)^2 \int
\frac{d^4 p}{(2\pi)^4} R\,
\gamma_\mu\,S^\beta_{\ell}(p)\gamma_\nu
\,L\,,
\label{w3}
\end{eqnarray}
in the above expressions $q=k-p$ as shown in Fig.~\ref{f:selfen1}. The
above equations contain the contribution to the neutrino self-energy
coming from the electrons and positrons in thermal equilibrium.  In
this article we do not consider the contribution to the neutrino
self-energy in vacuum with a magnetic field.

After integrating out the 3-momenta in the loop using the integration
results as presented in appendix \ref{app1} the form of Eq.~(\ref{w1})
becomes:
\begin{eqnarray}
\Sigma^W_1(k) &=& \frac{g^2}{M_W^2} \int_{-\infty}^\infty 
\frac{d p_0}{(2\pi)^4} \int_{-\infty}^\infty ds\,e^{\psi(p_0)}\,\tan z\, 
\eta_F(p_0)\nonumber\\
& & R \left[{\cal J}_0 {\cal R}_0 a_{30}(s) a_{20}(s') a_{10}(s')
- \frac{{\cal R}_0 a_{30}(s)}{M^2_W} \left\{a_{12}(s') a_{20}(s') + 
a_{22}(s') a_{10}(s')\right\}\right.\nonumber\\
&-&\left. \frac{{\cal R}_0}{M^2_W}
a_{32}(s)a_{20}(s')a_{10}(s') - \frac{2 a_{30}(s)}{M^2_W \sin z \cos z}
\left\{k_1 \gamma_1 a_{12}(s') a_{20}(s')+
k_2 \gamma_2 a_{22}(s')a_{10}(s')\right\}\right.\nonumber\\
&-& \left.\frac{2k_3}{M^2_W}\left\{\cot z \,{\rlap /b} + i{\rlap /u}\right\}
a_{32}(s)a_{20}(s')a_{10}(s')\right] L\,.
\label{wexp1}
\end{eqnarray}
The above expression has $a_{im}$s which are not tensor
components but Gaussian integrals defined as:
\begin{eqnarray}
a_{im}(s) \equiv \int_{-\infty}^{\infty} dp_i \, e^{-is p^2_i}\,p^m_i\,,
\label{aim}
\end{eqnarray}
where $i=1,\,2,\,3$ and $m=0,\,2$ and $s'=s\frac{\tan z}{z}$. In the
above equation $p^m_i$ is the $m$th power of the $i$th component of
$p$. The properties of this integral are briefly discussed in appendix
\ref{app1}. In the Eq.~(\ref{wexp1}),
\begin{eqnarray}
{\cal J}_0 &=& 1+\frac{p_0^2 - 2\omega_\ell p_0}{M_W^2}\,,
\label{j0}\\
{\cal R}_0 &=& p_0(\cot z \,{\rlap /u} + i{\rlap /b})\,.
\label{r0}
\end{eqnarray}
Using the form of the $a_{im}$s we can write Eq.~(\ref{wexp1}) as:
\begin{eqnarray}
\Sigma^W_1(k) =\pi^{3/2} e^{-3\pi i/4} \left(
\frac{e{\mathcal B}\,g^2}{M_W^2}\right) \int_{-\infty}^\infty 
\frac{d p_0}{(2\pi)^4} 
\int_{-\infty}^\infty
\frac{ds}{\sqrt{s}}\,e^{\psi(p_0)}\,\eta_F(p_0)\, {\cal W}(p_0, s)\,,
\label{wexp2} 
\end{eqnarray}
where,
\begin{eqnarray}
{\cal W}(p_0, s) &=& {\cal J}_0 p_0 \left(i {\rlap /b} + {\rlap /u} 
\cot z\right)
+ \frac{e {\mathcal B} p_0}{M_W^2}\left(\frac{i {\rlap /u}}{\sin^2 z}
 - {\rlap /b}  \cot z  - i  {\rlap /u}\right)\nonumber\\
&+& \frac{k_3}{s M_W^2}\left(i {\rlap /b}  \cot z - {\rlap /u}\right)
+ \frac{i e {\mathcal B}}{ M_W^2}\frac{(k_1 \gamma_1+ k_2 \gamma_2)}{\sin^2 z}
+ \frac{p_0}{2s M_W^2}\left( i {\rlap /u} \cot z -  {\rlap /b}\right)\,,
\label{wexp3}
\end{eqnarray}
Factors $\cot z$ and ${e{\mathcal B}}/{\sin^2 z}$ appearing in
the above expression can be written as:
\begin{eqnarray}
\cot z &=& i \sum_{n=0}^\infty \sum_{\lambda=\pm 1} e^{-is{\mathcal H}}\,,
\label{cot}\\
\frac{e {\mathcal B}}{\sin^2 z} &=& -\sum_{n=0}^\infty \sum_{\lambda=\pm 1}
{\mathcal H}e^{-is{\mathcal H}}\,,
\label{invsin2}
\end{eqnarray}
where,
\begin{eqnarray}
{\mathcal H}= e {\mathcal B}(2 n +1 -\lambda)\,. 
\label{hf}
\end{eqnarray}
Here $n$ is a positive integer including zero and $\lambda$ can take
only two values $\pm 1$ for $n\ne 0$ and for $n=0$, $\lambda=1$
only. The $n$ corresponds to the Landau level number occurring in the
energy of the charged leptons in a magnetic field and $\lambda$
corresponds to the spin states of the leptons.  Equations (\ref{cot})
and (\ref{invsin2}) are valid as long as ${\mathcal B}\ne 0$ as they
involve the landau quantization of the charged leptons in presence of
a magnetic field.  Using the above formulas for integrating the
parameter $s$ in Eq.~(\ref{wexp2}) and utilizing the fact that for
leptons in the magnetized medium \cite{Bhattacharya:2002qf}:
\begin{eqnarray}
p_0 \equiv E_{\ell, \,n} = \sqrt{m_\ell^2 + p_3^2 + {\mathcal H}}\,,
\label{ldisp}
\end{eqnarray}
and shifting the integration variable from $p_0$ to $p_3$ we get,
\begin{eqnarray}
\Sigma^W_1(k) &=& \frac{g^2}{4M_W^2}R\left[
\left(1 + \frac{m_\ell^2}{M_W^2}\right)\left\{(N^0_\ell - \bar{N}^0_\ell) 
{\rlap /b} + (N_\ell - \bar{N}_\ell){\rlap /u}\right\}\right.\nonumber\\
&-& \left.\frac{e {\mathcal B}}{M_W^2}\left\{(N_\ell - \bar{N}_\ell)
{\rlap /b} + (N^0_\ell - \bar{N}^0_\ell){\rlap /u}\right\}\right]L\nonumber\\
&-&  \frac{g^2 e {\mathcal B}}{M_W^4} \int_0^\infty \frac{dp_3}{(2\pi)^2}   
\sum_{n=0}^\infty \sum_{\lambda=\pm 1}
R\left[\left(\omega_\ell E_{\ell,\,n} \delta^{n,0}_{\lambda,1} +  
\frac{k_3 p_3^2}{E_{\ell,\,n}}\right){\rlap /b}\right.\nonumber\\
&+&\left.\left(\frac{k_3 p_3^2}{E_{\ell,\,n}}  \delta^{n,0}_{\lambda,1} +
\omega_\ell E_{\ell,\,n}\right){\rlap /u}
+ \frac{\mathcal H}{2E_{\ell,\,n}}{\rlap /k_\perp}\right]L(f_{\ell,\,n} +
 \bar{f}_{\ell,\,n})\,,
\label{wfin1}
\end{eqnarray}
where in the above expression,
\begin{eqnarray}
f_{\ell,\,n} = \frac{1}{e^{\beta(E_{\ell,\,n} - \mu_\ell)}+1}\,,\,\,&&
\bar{f}_{\ell,\,n} = \frac{1}{e^{\beta(E_{\ell,\,n} + \mu_\ell)}+1}\,,
\label{pdistrb}\\
N_\ell = \frac{e {\mathcal B}}{2\pi^2}\sum_{n=0}^\infty \sum_{\lambda=\pm 1}
\int_0^\infty dp_3 f_{\ell,\,n}\,,
\,\,&&
\bar{N}_\ell = \frac{e {\mathcal B}}{2\pi^2}\sum_{n=0}^\infty 
\sum_{\lambda=\pm 1}\int_0^\infty dp_3 
\bar{f}_{\ell,\,n}\,,
\label{nnbar}
\end{eqnarray}
and $N^0_\ell$ and $\bar{N}^0_\ell$ corresponds to $N_\ell$ and
$\bar{N}_\ell$ with $E_{\ell,\,n}$ in the distribution functions
replaced by $E_{\ell,\,0}$, that is $N^0_\ell$ and $\bar{N}^0_\ell$
are the particle and anti-particle number densities in the lowest
Landau level. The symbol $\delta^{n,0}_{\lambda,1}=1$ only when $n=0$
and $\lambda=1$ and zero in other cases.

Proceeding in exactly the same way as done in the previous analysis the 
form of $\Sigma^W_2(k)$ and $\Sigma^W_3(k)$ are given as:
\begin{eqnarray}
\Sigma^W_2(k) &=& \frac{g^2}{8M_W^4}R\left[(2k_3 {\rlap /k} - m_\ell^2 
{\rlap /b} + e {\mathcal B}{\rlap /u})(N^0_\ell - \bar{N}^0_\ell)
+ (2\omega_\ell {\rlap /k} + m_\ell^2{\rlap /u} + e {\mathcal B}{\rlap /b})
(N_\ell - \bar{N}_\ell)\right]L\nonumber\\
&+& \frac{g^2 e{\mathcal B}}{2M_W^4}
\int_0^\infty \frac{dp_3}{(2\pi)^2} \sum_{n=0}^\infty 
\sum_{\lambda=\pm 1} R\left[
\frac{\omega_\ell m_\ell^2}{E_{\ell,\,n}}\delta^{n,0}_{\lambda,1}\,\,
{\rlap /b}  - \frac{k_3 m_\ell^2}{E_{\ell,\,n}}\delta^{n,0}_{\lambda,1}\,\,
{\rlap /u} - \frac{m_\ell^2}{E_{\ell,\,n}}
{\rlap /k}\right]L (f_{\ell,\,n} + \bar{f}_{\ell,\,n})\,,
\label{wfin2}
\end{eqnarray}
and,
\begin{eqnarray}
\Sigma^W_3(k) = \frac{3g^2e{\mathcal B}}{8M_W^4}R\left[
(N^0_\ell - \bar{N}^0_\ell){\rlap /u} + (N_\ell - \bar{N}_\ell){\rlap /b})
\right]L\,.
\label{wfin3}
\end{eqnarray}
%
\subsection{The tadpole diagram}
\label{tadpd}
The tadpole contribution to the neutrino self-energy, up to one loop,
is given by Eq.~(\ref{tad}). 
Using the lepton propagator given by Eq.~(\ref{slbp}), which
corresponds to the magnetized plasma contribution, and the vacuum $Z$
boson propagator with zero momenta, the neutrino self-energy is:
\begin{eqnarray}
\Sigma^{T}(k) = \left(\frac{g}{2\cos\theta_{W}M_{Z}}\right)^2R\gamma^{\nu}
\int\frac{d^{4}p}{(2\pi)^{4}}
\int_{-\infty}^{\infty} ds\,\, {\rm Tr}\left[\gamma_{\nu}(c_{V}+c_{A}\gamma_{5})G(p,s)
\right]e^{\phi(p,s)}\eta_{F}(p_0)\,.
\label{tad1}
\end{eqnarray}
In this case also we only write that part of the neutrino self-energy
which arises from the electrons and positrons in the plasma and have
not written the vacuum contribution to the self-energy.

Using Eq.~(\ref{gps}) the trace is given by,
\begin{eqnarray}
{\rm Tr}\left[\gamma_{\nu}(c_{V}+c_{A}\gamma_{5})G(p,s)\right]=4\sec^2 z
(c_{V} A_{\nu}-ic_{A} B_{\nu})\,,
\label{trc}
\end{eqnarray}
where the 4-vectors $A_{\nu}$ and $ B_{\nu}$ are as given in
Eq.~(\ref{A}) and Eq.~(\ref{B}). Using the above equation the tadpole
contribution to the self-energy comes out as:
\begin{eqnarray}
\Sigma^{T}(k)=4\left(\frac{g}{2\cos\theta_{W}M_{Z}}\right)^2R\int
\frac{d^{4} p}{(2\pi)^{4}}
\int_{-\infty}^{\infty}ds\,\,e^{\phi(p,s)}\eta_{F}(p_0)
\sec^{2}z\,\,(c_{V} {\rlap A/}-ic_{A} 
{\rlap B/})\,.
\label{eq78}
\end{eqnarray}
Doing the $p_1$, $p_2$ and $p_3$ integrals, by using the results in 
appendix \ref{app1}, we obtain:
\begin{eqnarray}
\Sigma^{T}(k)&=&e{\mathcal B}\pi^{3/2}e^{-3\pi i/4}\left(\frac{g}{\cos\theta_{W}M_{Z}}
\right)^2R\int_{-\infty}^{\infty}\frac{p_0\,\,dp_{0}}{(2\pi)^4}
\int_{-\infty}^{\infty} \frac{ds}{\sqrt{s}} e^{\psi(p_0)}\eta_{F}(p_0)
\nonumber\\
&\times& \left[{\rlap /u}\,\,c_{V}\cot z - 
i {\rlap /b}\,\,c_{A}
\right]\,.
\end{eqnarray}
Using Eq.~(\ref{cot}) and doing some algebra the above equation can be 
written as:
\begin{eqnarray}
\Sigma^{T}(k)&=&e{\mathcal B}\pi^{3/2}e^{-3\pi i/4}
\left(\frac{g}{\cos\theta_W M_{Z}}\right)^2\int_{0}^
{\infty}\frac{p_0\,\,dp_{0}}{(2\pi)^4}
\int_{-\infty}^{\infty} 
\frac{ds}{\sqrt s} e^{\psi(p_0)}\nonumber\\
&& R \left(  ic_V {\rlap /u}\sum_{n=0}^\infty 
\sum_{\lambda =\pm 1} e^{-iS{\mathcal H}}- 
ic_A {\rlap /b}\right)L (f_{\ell,\,n}-\bar{f}_{\ell,\,n})\,,
\end{eqnarray}
where ${\mathcal H}$ is specified in Eq.~(\ref{hf}).  Now doing the
$s$ integral and re-introducing $p_3$ as the integration variable
through Eq.~(\ref{ldisp}) we obtain:
\begin{eqnarray}
\Sigma^{T}(k)&=&2e{\mathcal B}\pi^{2}\left(\frac{g}
{\cos\theta_{W}M_{Z}}\right)^2\int_{0}^
{\infty}\frac{dp_{3}}{(2\pi)^4} \sum_{n=0}^\infty \sum_{\lambda=\pm 1}
R \left(c_V {\rlap /u} - 
c_A \delta^{n,0}_{\lambda,1}\,\,{\rlap /b}\right) L(f_{\ell,\,n}-\bar{f}_{\ell,\,n})
\nonumber\\
&=&\left(\frac{g^2}
{4 \cos^2\theta_{W} M^2_{Z}}\right)R \left[c_V {\rlap /u}(N_\ell-\bar{N_\ell})-
c_A {\rlap /b}(N_\ell^0-\bar{N_\ell^0})\right]L\,,
\end{eqnarray}
where the symbols used in the above equation are explained in the last
subsection.
\subsection{The Z-exchange diagram}
\label{zexchd}
When the energy of the neutrinos is much greater than their chemical
potentials, the result of the $Z$ exchange process is known
\cite{BravoGarcia:2007mv}:
\begin{eqnarray}
\Sigma^{Z}(k) &=& \left(\frac{g}{2 \cos \theta_W M^2_Z}\right)^2 R
\left[\left\{\omega_\ell (N_{\nu_\ell} - \bar{N}_{\nu_\ell}) + 
\frac{2}{3}(\langle \omega_\ell^{\rm B}\rangle N_{\nu_\ell} + \langle 
{\bar{\omega}^{\rm B}_\ell}\rangle\bar{N}_{\nu_\ell})\right\} 
{\rlap /k}\right.\nonumber\\
&+& \left.\left(\frac{g}{2 \cos \theta_W M_Z}\right)^2
\left\{(N_{\nu_\ell} - \bar{N}_{\nu_\ell}) - \frac{8}{3}\frac{\omega_\ell}
{M_Z^2}(\langle \omega_\ell^{\rm B}\rangle N_{\nu_\ell} + \langle 
{\bar{\omega}^{\rm B}_\ell}\rangle
\bar{N}_{\nu_\ell})\right\} {\rlap /u}\right] L\,,
\label{zexpr}
\end{eqnarray}
where,
\begin{eqnarray}
N_{\nu_\ell} = \int \frac{d^3 p}{(2\pi)^3} \frac{1}
{e^{\beta(\omega_\ell^{\rm B} - 
\mu_{\nu_\ell})}+1}\,,\,\,\,
{\bar N}_{\nu_\ell} = \int \frac{d^3 p}{(2\pi)^3} \frac{1}
{e^{\beta(\omega_\ell^{\rm B} + 
\mu_{\nu_\ell})}+1}\,,
\label{nunom}
\end{eqnarray}
are the number densities of the neutrinos and antineutrinos. Here
$\omega_\ell^{\rm B}$ is the energy of the background
neutrinos and $\langle \omega_\ell^{\rm B}\rangle$ 
($\langle {\bar{\omega}^{\rm B}_\ell}\rangle$) is the average 
neutrino (anti-neutrino) energy per unit volume per neutrino (anti-neutrino).

\section{The dispersion relation}
\label{dispr}
We can write the forms of $b$ and $c$ from the discussions on the last
section and from Eqs.~(\ref{at}), (\ref{bt}) and (\ref{ct}). The forms
of $b$ and $c$ are:
\begin{eqnarray}
b &=&  \frac{g^2 (N_\ell-\bar{N_\ell})}{4M_W^2}\left(1 + c_V + 
\frac{3 m_\ell^2}{2 M_W^2}\right)
+\frac{g^2 e {\mathcal B}}{4M_W^4}(N^0_\ell - \bar{N}^0_\ell)  \nonumber\\
&+& \frac{g^2}{4M_W^2}
\left[(N_{\nu_\ell} - \bar{N}_{\nu_\ell}) - \frac{8}{3}\frac{\omega_\ell}
{M_Z^2}
(\langle \omega_\ell^{\rm B}\rangle N_{\nu_\ell} + 
\langle {\bar{\omega}^{\rm B}_\ell} \rangle
\bar{N}_{\nu_\ell})\right]\nonumber\\
&-&\frac{e {\mathcal B} g^2}{M_W^4} \int_0^\infty \frac{dp_3}{(2\pi)^2}
\sum_{n=0}^\infty \sum_{\lambda=\pm 1}
\left[\frac{k_3}{E_{\ell,\,n}}\left( p_3^2  + 
\frac{m_\ell^2}{2}\right)\delta^{n,0}_{\lambda,1} +
\omega_\ell E_{\ell,\,n}\right](f_{\ell,\,n} + \bar{f}_{\ell,\,n})\,,
\label{bcompl}\\
c &=&  \frac{g^2 (N^0_\ell-\bar{N}^0_\ell)}{4M_W^2}\left(1 - c_A +
\frac{m_\ell^2}{2M_W^2}\right)
+\frac{g^2 e {\mathcal B}}{4M_W^4}(N_\ell - \bar{N}_\ell)  
\nonumber\\
&-&\frac{e {\mathcal B} g^2}{M_W^4} \int_0^\infty \frac{dp_3}{(2\pi)^2}
\sum_{n=0}^\infty \sum_{\lambda=\pm 1}
\left[\omega_\ell\left(E_{\ell,\,n} - \frac{m_\ell^2}{E_{\ell,\,n}}\right)
\delta^{n,0}_{\lambda,1} + 
\frac{k_3 p_3^2}{E_{\ell,\,n}}\right](f_{\ell,\,n} + \bar{f}_{\ell,\,n}) \,.
\label{ccompl}
\end{eqnarray}
In our calculation we see that in general we cannot say a-priori that
the magnitude of $a_\parallel$ or $a_\perp$ is smaller than the other
coefficients in the self-energy expression. Consequently we specify
the forms of $a_\parallel$ and $a_\perp$:
\begin{eqnarray}
a_\perp &=& -\frac{g^2 e {\mathcal B}}{M_W^4} \int_0^\infty \frac{dp_3}
{(2\pi)^2}\sum_{n=0}^\infty \sum_{\lambda=\pm 1} \left(\frac{\mathcal H}
{2E_{\ell,\,n}}+ \frac{m_\ell^2}{E_{\ell,\,n}}\right)(f_{\ell,\,n} 
+ \bar{f}_{\ell,\,n})\nonumber\\
&+& \frac{g^2}{4M_W^4}\left[k_3(N^0_\ell - \bar{N}^0_\ell) + 
\omega_\ell (N_{_\ell} - \bar{N}_{_\ell} + N_{\nu_\ell} - \bar{N}_{\nu_\ell})
+ \frac{2}{3}(\langle \omega_\ell^{\rm B}\rangle N_{\nu_\ell} + \langle
{\bar{\omega}^{\rm B}_\ell}\rangle\bar{N}_{\nu_\ell})\right]\,,
\label{aperpf}\\
a_\parallel &=&  -\frac{g^2 e {\mathcal B}}{M_W^4} \int_0^\infty 
\frac{dp_3}{(2\pi)^2}
\sum_{n=0}^\infty \sum_{\lambda=\pm 1} \frac{m_\ell^2}{E_{\ell,\,n}}
(f_{\ell,\,n} + 
\bar{f}_{\ell,\,n})\nonumber\\
&+& \frac{g^2}{4M_W^4}\left[k_3(N^0_\ell - \bar{N}^0_\ell) +
\omega_\ell(N_{_\ell} - \bar{N}_{_\ell} + N_{\nu_\ell} - \bar{N}_{\nu_\ell})
+ \frac{2}{3}(\langle \omega_\ell^{\rm B}\rangle N_{\nu_\ell} + \langle
{\bar{\omega}^{\rm B}_\ell}\rangle\bar{N}_{\nu_\ell})\right]\,.
\label{aparaf}
\end{eqnarray}
The above forms of the coefficients determine the dispersion relation
of the neutrino in a magnetized plasma containing charged leptons. In
writing the above equations we have used $M_W=M_Z \cos \theta_W$.
Equations (\ref{bcompl}), (\ref{ccompl}), (\ref{aperpf}) and
(\ref{aparaf}) are the central results of this article.  The
expressions of $a_\parallel$ and $a_\perp$ only has the leptonic
contributions as we did not take into account the Landau quantization
of the charged gauge fields. If we used the full $W$ propagator
instead of the one in Eq.~(\ref{impw}), containing only the linear
order correction $e{\mathcal B}/ M^4_W$ to the $W$ propagator, then
$a_\parallel$ and $a_\perp$ would get contributions from the gauge
sector also. In the absence of the Landau quantization of the charged
gauge fields it is noticed that $a_\parallel-a_\perp$ diminishes as
the magnetic field increases and the temperature of the system remains
constant.  The coefficient $b$ closely resembles the corresponding
coefficient in \cite{D'Olivo:1989cr} in the four-Fermi limit.  Both
$b$ and $c$ match exactly with the corresponding forms in the
four-Fermi limit with the results obtained in \cite{Erdas:1998uu}. But
the expressions of $a_\perp$ and $a_\parallel$ in Equations
(\ref{aperpf}) and (\ref{aparaf}) has not been calculated before, in
general. Previously $a_\parallel-a_\perp$ has only been calculated in
the {\bf CP} symmetric plasma but it must be noted that
$a_\parallel-a_\perp$ exists in the {\bf CP} asymmetric plasma
also. 

Of special interest is the charge symmetric plasma, which perhaps
existed in the early universe. In this situation where the chemical
potentials of the particles are negligible, the coefficients become,
\begin{eqnarray}
b &=&  -\frac{4 g^2 \omega_\ell}{3 M_W^2 M_Z^2}  
\langle \omega_\ell^{\rm B}\rangle N_{\nu_\ell}\nonumber\\ 
&-& \frac{2 e {\mathcal B}
g^2}{M_W^4} 
\int_0^\infty \frac{dp_3}{(2\pi)^2}
\sum_{n=0}^\infty \sum_{\lambda=\pm 1}
\left[\frac{k_3}{E_{\ell,\,n}}\left( p_3^2  +
\frac{m_\ell^2}{2}\right)\delta^{n,0}_{\lambda,1} +
\omega_\ell E_{\ell,\,n}\right]f_{\ell,\,n}\,,
\label{bcompls}\\
c &=& - \frac{2e {\mathcal B} g^2}{M_W^4} \int_0^\infty \frac{dp_3}{(2\pi)^2}
\sum_{n=0}^\infty \sum_{\lambda=\pm 1}
\left[\omega_\ell\left(E_{\ell,\,n} - \frac{m_\ell^2}{E_{\ell,\,n}}\right)
\delta^{n,0}_{\lambda,1} +
\frac{k_3 p_3^2}{E_{\ell,\,n}}\right]f_{\ell,\,n} \,.
\label{ccompls}\\
a_\perp &=& -\frac{2 g^2 e {\mathcal B}}{M_W^4} \int_0^\infty 
\frac{dp_3}{(2\pi)^2}
\sum_{n=0}^\infty \sum_{\lambda=\pm 1} \left(\frac{\mathcal H}{2E_{\ell,\,n}}
+ \frac{m_\ell^2}{E_{\ell,\,n}}\right) f_{\ell,\,n}
+\frac{g^2}{3M_W^4}\langle \omega_\ell^{\rm B}\rangle N_{\nu_\ell}\,,
\label{aperpfs}\\
a_\parallel &=&  -\frac{2 g^2 e {\mathcal B}}{M_W^4} \int_0^\infty 
\frac{dp_3}{(2\pi)^2}
\sum_{n=0}^\infty \sum_{\lambda=\pm 1} \frac{m_\ell^2}{E_{\ell,\,n}}
f_{\ell,\,n}
+\frac{g^2}{3M_W^4}\langle \omega_\ell^{\rm B}\rangle N_{\nu_\ell}\,.
\label{aparafs}
\end{eqnarray}
>From the above expressions we immediately notice that all the
contributions in the charge symmetric case are proportional to
$M_W^{-4}$ or $G_F^2$. From the above equations
if we neglect all the terms explicitly containing the mass of the
charged leptons then we reproduce closely the form of the
neutrino-self energy calculated in \cite{Erdas:1998uu} in a {\bf CP}
symmetric plasma. 

In Eq.~(\ref{mndisp}) the most general form of the
neutrino dispersion relation was written. To order of $g^2$ the
dispersion relation becomes,
\begin{eqnarray}
\omega_{\ell} = \left[|{\bf k}|^2 - 2 c {\bf k}\cdot {\hat{\bf b}} 
+ 2(a_\parallel - a_\perp) k_\perp^2 \right]^{1/2} +b\,,
\end{eqnarray}
where we have taken the positive sign of the square root in
Eq.~(\ref{mndisp}).  The above dispersion relation can be simplified
by binomially expanding the square root and neglecting terms of order
more than $g^2$. The expansion gives,
\begin{eqnarray}
\omega_{\ell} &=& |{\bf k}| -  c {\hat{\bf k}}\cdot {\hat{\bf b}}
+ (a_\parallel - a_\perp) \frac{k_\perp^2}{|{\bf k}|} +b\,,\nonumber\\
&=& |{\bf k}| -  c \cos \theta 
+ (a_\parallel - a_\perp) {|{\bf k}|} \sin^2 \theta +b\,,
\label{fdispr}
\end{eqnarray}
where $k^3 = k_z = |{\bf k}| \cos \theta$. The above equation implies
that in presence of a magnetized medium the effective-potential acting
on the neutrinos is of the form,
\begin{eqnarray}
V_{\rm eff} = b -  c \cos \theta
+ (a_\parallel - a_\perp) {|{\bf k}|} \sin^2 \theta \,.
\label{veff}
\end{eqnarray}
>From the expressions of $a_\parallel$ and $a_\perp$ in the {\bf CP}
symmetric case we see that in the lowest Landau level $a_\parallel -
a_\perp$ is zero and in that case the effective potential is
independent of $a$. With the form of the effective potential in
Eq.~(\ref{veff}) the problem of neutrino oscillations in the {\bf CP}
symmetric magnetized plasma in the early universe can be tackled.

The calculation of the coefficients $c$, $(a_\parallel-a_\perp)$,
which are directly related to the presence of a non-zero magnetic
field, do not allow us to take the ${\mathcal B}\to 0$ limit because
of the non-perturbative nature of the Landau quantization of the
charged fermions.  Interestingly it turns out that in the zero external
magnetic field limit the form of $b$ exactly matches the form in an
unmagnetized plasma.  We can write Eq.~(\ref{bcompls}) as:
\begin{eqnarray}
b  &=&  - \frac{4 g^2 \omega_\ell}{3 M_W^2 M_Z^2}  
\langle \omega_\ell^{\rm B}\rangle N_{\nu_\ell} - 
\frac{e {\mathcal B} g^2 \omega_\ell}{M_W^4} \int_0^\infty \frac{dp_3}{2\pi^2}
E_{\ell,\,n}\,f_{\ell,\,n} \nonumber\\
&-& \frac{2 e {\mathcal B} g^2}{M_W^4} \int_0^\infty \frac{dp_3}{(2\pi)^2}
\sum_{n=0}^\infty \sum_{\lambda=\pm 1}
\left[\frac{k_3}{E_{\ell,\,n}}\left( p_3^2  +
\frac{m_\ell^2}{2}\right)\delta^{n,0}_{\lambda,1}\right]f_{\ell,\,n}\,.
\end{eqnarray}
Now if we substitute the lepton energy in the loop integral by the thermal
average of the lepton energy and use the relation  
$\bar{N}_\ell = \frac{e {\mathcal B}}{2\pi^2}\sum_{n=0}^\infty 
\sum_{\lambda=\pm 1}\int_0^\infty dp_3 f_{\ell,\,n}$, the above equation 
becomes,
\begin{eqnarray}
b &=&  - \frac{16\sqrt{2} G_F \omega_\ell}{3 M_Z^2}  
\langle \omega_\ell^{\rm B}\rangle N_{\nu_\ell} - 
\frac{16 \sqrt{2} G_F \omega_\ell}{M_W^2} 
\langle E_{\ell} \rangle \,N_\ell\nonumber\\ 
&-& \frac{2 e {\mathcal B} g^2}{M_W^4} \int_0^\infty \frac{dp_3}{(2\pi)^2}
\sum_{n=0}^\infty \sum_{\lambda=\pm 1}
\left[\frac{k_3}{E_{\ell,\,n}}\left( p_3^2  +
\frac{m_\ell^2}{2}\right)\delta^{n,0}_{\lambda,1}\right]f_{\ell,\,n}\,,
\label{bzero}
\end{eqnarray}
where we have utilized $\sqrt{2} G_F = \frac{g^2}{4 M_W^2}$ and
replaced $E_{\ell, n}$ by $4/3 \langle E_{\ell} \rangle$ because there are
three flavours and two spin states for the electron and the
positron.  In the ${\mathcal B} \to 0$ limit the last term on the right
hand side of the above equation vanishes and we have,
\begin{eqnarray}
b_{{\mathcal B}\to 0}=- \frac{16\sqrt{2} G_F \omega_\ell}{3 M_Z^2}  
\left(\langle E_{\nu^{\rm B}_\ell}\rangle N_{\nu_\ell} -  
\langle E_{\ell} \rangle \,N_\ell\right)\,.
\label{nr}
\end{eqnarray}
The Eq.~(\ref{nr}) resembles the results found in \cite{notr}. The
signs are different as in \cite{notr} the authors used the opposite
sign for $b$.
\section{Conclusion}
\label{conclu}
In the present work we calculated the self-energy of the neutrino in a
medium seeded with a uniform classical magnetic field. The
calculations were carried out in the unitary gauge where the
unphysical Higgs contribution does not appear. The background is
supposed to be comprised of the charged leptons and neutrinos
equillibriated at the same temperature. The magnitude of the magnetic
field is such that only linear contributions of the field appear in
the charged $W^{\pm}$ boson propagators but all orders of the field
are present in the charged lepton propagators. Equations
(\ref{bcompl}), (\ref{ccompl}), (\ref{aperpf}) and (\ref{aparaf})
contain the most general results of the self-energy calculation in the
absence of Landau quantization of the charged gauge fields. Earlier
calculations have been done in this regard but the authors have used
separate assumptions to calculate the expression of the neutrino
self-energy in various cases. The expression of the neutrino
self-energy in various specific cases can be calculated from the
general results given in the previous section. The correspondence of
our result with results from \cite{Erdas:1998uu} confirms the gauge
invariance of the calculations as in the last reference the authors
used a different gauge to calculate the neutrino-self energy. We point
out the special role played by the Landau quantization of the charged
leptons in the medium in modifying the neutrino self-energy through
the coefficient $a_\parallel-a_\perp$.  The zero field limit of the
coefficient $b$ matches perfectly with the form predicted in presence
of an unmagnetized medium.

The main application of our results are expected to be in the early
universe, specifically before the period of neutrino decoupling which
was expected to happen around an energy scale of $1\,{\rm MeV}$ when
the age of the universe was about 1 second. After that period the
neutrinos decoupled from the thermal equilibrium and its temperature
started to red-shift with time. Other possible application of our
calculation can be in GRB fireballs \cite{Sahu:2005zh} where neutrinos
propagate in the presence of a electron, positron plasma with a very
small baryon contamination.
\vskip .2cm
\noindent{\bf Acknowledgement}: This research is partially supported
by DGAPA-UNAM project IN119405 and Conacyt, Mexico, grant No. 52975.
\appendix\section*{\hfil Appendix \hfil}
\section{Gaussian Integral results}
\label{app1}
The integral in Eq.~(\ref{aim}) is:
\begin{eqnarray}
a_{im}(s) = \int_{-\infty}^{\infty} dp_i \, e^{-is p^2_i}\,p^m_i\,,
\end{eqnarray}
where $i=1,\,2,\,3$ and $m=0,\,2$.  In the actual calculations the
$p_\perp$ components, which stands for $(p_1,\,p_2)$, and the $p_3$
component are integrated in different manner as for the perpendicular
components we have
\begin{eqnarray}
a_{im}(s') = \int_{-\infty}^{\infty} dp_i \, 
e^{-is'p^2_i}\,p^m_i\,,
\end{eqnarray}
where $s'=s\frac{\tan z}{z}$. In practice $a_{1m}$ and $a_{2m}$ are different 
from $a_{3m}$. In general we will have,
\begin{eqnarray}
a_{10}(s') &=& a_{20}(s') = \sqrt{\frac{\pi}{s,}} 
\exp[\frac{-i\pi}{4}]=\sqrt{\frac{\pi z}{s\tan z}}\exp[\frac{-i\pi}{4}]\,,\\
a_{12}(s') &=& a_{22}(s') = \frac{\sqrt{\pi}}{2 s'^{3/2}}\exp[
\frac{-3\pi i}{4}]=\frac{\sqrt{\pi}}{2}\left(\frac{z}{s\tan z}\right)^{3/2}
\exp[\frac{-3\pi i}{4}]\,,
\end{eqnarray}
and,
\begin{eqnarray}
a_{30}(s) = \sqrt{\frac{\pi}{s}}
\exp[\frac{-i\pi}{4}]\,,\,\,
a_{32}(s) = \frac{\sqrt{\pi}}{2 s^{3/2}}\exp[
\frac{-3\pi i}{4}]\,.
\end{eqnarray}
>From the above results it can be shown, 
\begin{eqnarray}
a_{32} (s) =-\frac{i}{2s}\,a_{30}(s)\,,\,\,
a_{12}(s') = -\frac{iz}{2s\tan z}\,a_{10}\,,\,\,
a_{22}(s') = -\frac{iz}{2s\tan z}\,a_{20}\,.
\end{eqnarray}
%

\end{document}